\documentclass[a4paper,11pt]{article}
\usepackage[super,compress]{cite}
\usepackage{graphicx}
\usepackage{dcolumn}
\usepackage{bm}
\usepackage{epsfig,amsmath,graphicx,amssymb,amsmath,amsfonts,epsfig}
\usepackage[usenames,dvipsnames]{color}
\usepackage{appendix}
\usepackage{textcomp}
\usepackage{float}
\usepackage{subfigure}
\title{Bekenstein-Hawking Entropy Products for NUT class of Black Holes in AdS Space}
\author{Parthapratim Pradhan\footnote{pppradhan77@gmail.com}\\ 
\textit{Department of Physics}\\
\textit{Hiralal Mazumdar Memorial College For Women}\\
{Dakshineswar, Kolkata-700035, India}}
\date{\today}

\begin{document}

\maketitle

\begin{abstract}
We derive the entropy product rule for Taub-NUT~(Newman-Unti-Tamburino)-de~Sitter black hole~(BH)
and Taub-NUT--Anti-de~Sitter BH. We show that the entropy products in terms of both the physical 
horizons  are \emph{mass-independent}. Both \emph{perturbative} approximation and  \emph{direct} 
method have been considered. By introducing the cosmological horizon we show that for 
Taub-NUT-de~Sitter BH, there exists a mass-independent entropy functional relation in terms of 
three horizons namely event horizon~(EH), Cauchy horizon~(CH) and cosmological horizon~(CHH) which 
depends on cosmological parameter~($\Lambda$) and the NUT parameter~($N$). 
For Taub-NUT-anti-de~Sitter BHs, we determine the mass-independent entropy functional 
relations in terms of two physical horizons~(namely EH and CH) which depends on only NUT parameter. 
Some-times some complicated functions of EH entropy and CH entropy are also strictly mass-independent. 
This is plausible only due to the new formalism developed in~\cite{wu}~[Phys. Rev. D 100, 101501(R)~(2019)] 
for NUT class of BHs. The formalism states that a generic four dimensional Taub-NUT spacetime should 
be described completely 
in terms of three or four different types of thermodynamic hairs. They could be defined as the Komar 
mass~($M=m$), the angular momentum~($J_{n}=mn$), the gravitomagnetic charge ($N=n$), the dual~(magnetic) 
mass $(\tilde{M}=n)$. Finally, we could  say that this universality is mainly  due to the presence of 
\emph{new conserved charges  $J_{N}=MN$} which is closely analogue  to the  Kerr like angular 
momentum $J=aM$.
\end{abstract}

\newpage

\textheight 25 cm

\clearpage


\section{Introduction}
Area~(or entropy) products~\cite{visser13,ansorg09,cvetic10,pope14,pp15,meng14,plb20,grg21,jetp} are 
very fascinating topic of research in BH physics. 
These products are constructed in terms of the entropy of event~(outer) horizons~(${\cal H}^{+}$) and 
Cauchy~(inner) horizons~(${\cal H}^{-}$). Again these horizons are called as physical horizons
\footnote{The fact that CH is a blue-shift region while EH is a red-shift region by its own virtue. 
It should be noted that  the CH is highly unstable due to the exterior perturbation~\cite{sc83}}. The 
products are fascinating in a sense that they are independent of the 
 Arnowitt-Deser-Misner~(ADM) mass of the BH spacetime and depends only on 
various quantised charges, \emph{quantised} angular momentum and 
moduli etc.. For example, it should be noted that for most general axisymmetric
Kerr-Newman class of BH having EH and CH, the entropy product of ${\cal H}^{\pm}$ should be 
\begin{eqnarray}
S_{+} S_{-} &=& 4\pi^2 \left[J^2+\frac{ Q^4}{4}\right] ~.\label{TNp}
\end{eqnarray}
The fundamental physics is that due to stationarity there is an absence of  
matter between the two horizons~\cite{visser13}.
Now a vital question is that what is the origin of Bekenstein-Hawking~\cite{bk73,bcw73} 
entropy $S_{\pm}=\frac{A_{\pm}}{4}$ at the microscopic level?  This is a real problem in quantum theory 
of gravity. Here $A_{\pm}$ is the area of the black hole EH and CH. There has been a momentous 
success in four and five dimensions for super-symmetric Bogomol'ni-Prasad-Sommerfield~(BPS) 
class of  black holes~\cite{vafa}. 

Where it was exhibited that the microscopic degrees of 
freedom could be explained in terms of 2D conformal-field-theory~(CFT). If there exists 
BH entropy~($S_{+}$) for EH then it is also equally authenticate that there exists 
BH entropy~($S_{-}$) for CH. Hence there might be a relevance of inner horizon (CH) in BH mechanics to 
understanding the microscopic nature of inner BH entropy in compared with the outer BH entropy.
For extremal Kerr spacetime, the microscopic entropy should 
be found in ~\cite{guica}. 
However, the detailed information of the microscopic entropy 
of non-extremal BH still  to be obscure to us. Some fruitful progress have been indicated
in~\cite{castro10}. 
For  BPS category of BHs, the entropy product of ${\cal H}^{\pm}$ is derived 
to be~\cite{cvetic10}
\begin{eqnarray}
S_{+} S_{-}  &=& (2\pi)^2 \left[\sqrt{N_{L}}+\sqrt{N_{R}}\right] \left[\sqrt{N_{L}}-\sqrt{N_{R}}\right]\\
             &=& 4\pi^2 N , \,\, N\in {\mathbb{N}}, N_{L}\in {\mathbb{N}}, N_{R} \in {\mathbb{N}} ~.\label{tnl}
\end{eqnarray} 
The integers $N_{L}$ and $N_{R}$ can be interpreted as the excitation numbers of 
the left and right moving modes of a weakly-coupled 2D CFT.  The parameters $N_{L}$ and 
$N_{R}$ are depends explicitly on all the BH parameters.  
This indicates that the product should  be quantized~\cite{finn97,cv96,cv97,cvf97} 
and it is an integer quantity.

Sometimes the conjectured \emph{mass-independence} also breaks down. For instance, by 
introducing the cosmological constant, it has been shown that the entropy products for 
Schwarzschild-de-Sitter~(Kottler) BH in terms of EH and CH are \emph{not} mass-independent~\cite{visser13}. 
By adding the third horizon namely virtual horizon~(VH)~(even though unphysical), the results do not change. 
For Schwarzschild-anti-de-Sitter BHs, there exists three horizons: one physical horizon~(called as EH)
and other two horizons are VH~(unphysical). Hence the results indicates 
the entropy products in terms of these horizons are \emph{not} mass-independent. 

Introducing the charge parameter in Schwarzschild-de-Sitter spacetime 
i.e. for Reissner-Nordstr\"{o}m-de~Sitter BH, the situation is  
little bit improved. Here, the Killing horizons equations are fourth order 
polynomial. There exists three physical horizons~(EH, CH and CHH) and one VH.  
So, the entropy products in terms of three physical horizons are 
mass-independent~\cite{visser13}. 
Though they are some complicated function of physical horizons but they are no less 
than mass-independent. Similarly, for Reissner-Nordstr\"{o}m-anti-de~Sitter  BH, 
there exists two physical horizons and two VHs. So, the entropy products are determined 
perturbatively and it was shown that the products are not mass-independent. 
In a nutshell, the entropy products are \emph{not generic}.

However, in this work  interestingly we  prove the entropy products for NUT class of BHs in AdS 
space have \emph{generic} feature. Specifically, we have considered \emph{Taub-NUT-de~Sitter BHs} and 
\emph{Taub-NUT--anti-de~Sitter} BHs. First, we have calculated perturbatively the entropy products in 
terms of EH and CH. We prove that the products are \emph{mass-independent}. Next we compute 
the entropy products in terms of three physical horizons by direct method and show that the 
functional form of three physical horizons is independent of the mass parameter but it 
depends on only the NUT parameter and cosmological constant. 
Next we calculate the mass-independent relation in terms of two 
physical horizons. 

In some cases, there exists mass-dependent relation in terms of two physical horizons but 
introducing the formalism developed in~\cite{wu}~[See also \cite{wu22}] we can reformulate
it mass-independent form.
The formalism argued that a generic four dimensional Taub-NUT spacetime can be described completely 
in terms of three or four different types of thermodynamic hairs. They should be defined as the 
Komar mass~($M=m$), the angular momentum~($J_{n}=mn$), the gravitomagnetic charge ($N=n$), 
the dual~(magnetic) mass $(\tilde{M}=n)$. The generic feature comes solely  due to the 
presence of new conserved charges  $J_{N}=MN$ which is closely analogue to the  Kerr like 
angular momentum $J=aM$.

In our previous work, we have already examined the area (or entropy) products for NUT 
class of BHs~\cite{plb20}. We have considered there the Lorentzian Taub-NUT BH, Euclidean 
Taub-NUT BH, Reissner-Nordstr\"{o}m--Taub-NUT, Kerr-Taub-NUT BH and  Kerr-Newman-Taub-NUT BH. 
In each cases, we proved that the entropy product of both physical horizons is universal in 
nature after the incorporation of angular momentum like parameter $J_{N}=MN$ with the generic 
parameters $M$ and $N$ as a global conserved charges. However we have \emph{not} considerd 
the AdS spacetime. Here we have explicitly considered the 
\emph{AdS spacetime particularly Taub-NUT-de~Sitter BHs and Taub-NUT-anti-de~Sitter BHs}. 
This is an another strong reason to investigate this work.

It should be mentioned that in recent years various aspects of thermodynamics of Lorentzian 
Taub-NUT BH in~($3+1$) dimensional Einstein-Maxwell gravity have studied by many authors
in different perspectives. We should mention some of good references~
\cite{chandra18,chandra19}
[See also\cite{bn69,manko05,dk74,as,hunter98,hunter99,page99,myers99,empa99}] 
for our readers.

The paper is organized as follows. In the next section, we will consider the Taub-NUT-de~Sitter 
BHs and examined the mass-independence feature both perturbatively and direct method. 
Analogously in Sec.~(\ref{tnads}), we will examine the mass-independent feature  both perturbatively 
and direct method for Taub-NUT--anti-de~Sitter BHs.  Finally, in Sec.~(\ref{con}), we have given 
the conclusions of the work.

\section{\label{tnds}~Taub-NUT-de~Sitter BHs}
The Taub-NUT--de-Sitter spacetime in Schwarzschild like coordinates has the 
form~\cite{taub,mkg,mis,miller,griffith}[Page-236, Griffiths \& Podolsk\'{y}]
\begin{eqnarray}
ds^2 &=& -\Upsilon(r) \, \left(dt+2n\cos\theta d\phi\right)^2+ \frac{dr^2}{\Upsilon(r)}+\left(r^2+n^2\right) \left(d\theta^2
+\sin^2\theta d\phi^2 \right) ~,\label{tn}
\end{eqnarray}
where the function $\Upsilon(r)$ is given by 
\begin{eqnarray}
\Upsilon(r) &=& \frac{1}{r^2+n^2} \left[r^2-n^2-2mr-\frac{\Lambda}{3}\left(r^4+6n^2r^2-3n^4\right) \right]
\end{eqnarray}
We know the global conserved charges for Taub-NUT spacetime~\cite{wu} is 
\begin{eqnarray}
m &=& M~\mbox{(Komar mass)} \nonumber\\
n &=& N ~\mbox{(Gravitomagnetic charge)}\nonumber\\
m\,n &=& J_{n}=J_{N}=M\,N~ \mbox{(Angular momentum)} ~.\label{tnn}
\end{eqnarray}

If we incorporate the condition [Eq.~(\ref{tnn})], then we may rewrite the metric as 
\begin{eqnarray}
ds^2 &=& -\Upsilon(r) \, \left(dt+2N\cos\theta d\phi\right)^2+ \frac{dr^2}{\Upsilon(r)}+\left(r^2+N^2\right) \left(d\theta^2
+\sin^2\theta d\phi^2 \right) ~,\label{tn1}
\end{eqnarray}
and the function $\Upsilon(r)$ is defined by  
\begin{eqnarray}
\Upsilon(r) &=& \frac{1}{r^2+N^2} \left[r^2-N^2-2Mr-\frac{\Lambda}{3}\left(r^4+6N^2r^2-3N^4\right) \right]
\end{eqnarray}
It may be noted that the metric reduces to Schwarzschild-de~Sitter~(Kottler) spacetime when $N=0$. 
The Hawking temperature of the BH is
$$
T_{\pm}=\frac{{\kappa}_{\pm}} {2\pi}=\frac{\Upsilon'(r_{\pm})}{4\pi \left(r_{\pm}^2+N^2\right)}
$$
where $\kappa_{\pm}$ is the surface gravity of the BH. The first law of thermodynamics 
with cosmological constant is derived in Ref.~\cite{wu} and without cosmological constant is derived 
in \cite{grg21}.

Now the Killing horizons are determined by solving the following equation
\begin{eqnarray}
\Upsilon(r) &=& 0
\end{eqnarray}
This can be rewritten as 
\begin{eqnarray}
r^2-N^2-2Mr-\frac{\Lambda}{3}\left(r^4+6N^2r^2-3N^4\right) &=& 0~\label{eq1.1}
\end{eqnarray}
For $\Lambda>0$, it is advantageous to put 
$$
b^2=\frac{1}{\Lambda},
$$
where $b$ denotes the spatial radius of curvature. Thus 
\begin{eqnarray}
 r^4-3 (b^2-2N^2) r^2+6M b^2r+3b^2N^2-3N^4 &=& 0  ~\label{eq1.2}
\end{eqnarray}
In the limit $\Lambda\rightarrow 0$ (or $b\rightarrow \infty$) we find standard Taub-NUT spacetime. 
Also $N\rightarrow 0$ and $\Lambda \neq 0$ gives  Schwarzschild--de~Sitter~(Kottler) spacetime. The 
quartic equation~[Eq.~(\ref{eq1.2})] can be rewritten as 
\begin{eqnarray}
 r^4-3 (b^2-2N^2) \left[r^2-2M \left(\frac{b^2}{b^2-2N^2}\right)r
 +\left(\frac{N^4-b^2N^2}{b^2-2N^2} \right)\right] &=& 0  ~\label{eq1.3}
\end{eqnarray}
It can be {reformulated} as 
\begin{eqnarray}
 r^4-3 (b^2-2N^2) (r-r_{+})(r-r_{-}) &=& 0  ~\label{eq1.4}
\end{eqnarray}
where 
$$
r_{+}+r_{-}=\frac{2M}{\left(1-2\frac{N^2}{b^2}\right)}
$$
and 
$$
r_{+}\,r_{-}=\frac{N^2\left(\frac{N^2}{b^2}-1\right)}{\left(1-2\frac{N^2}{b^2}\right)}
$$
where $r_{\pm}$ denote two roots of the term of the square bracket in Eq.~(\ref{eq1.3}), and 
they reduces to the horizon radii of the Taub-NUT solution when $b\rightarrow \infty$. 

\subsection{Perturbative Analysis}
The quartic equation in Eq.~(\ref{eq1.4}) has an exact solution but it is more convenient to find 
the roots perturbatively. To do that we can re-write the above equation as 
\begin{eqnarray}
r  &=&  r_{\pm} + \frac{r^4}{3\left(b^2-2N^2 \right) \left(r-r_{\mp} \right)}.~\label{eq1.5} 
\end{eqnarray}
Now we can solve it perturbatively. 

\subsubsection{Event and Cauchy Horizons}
To a first approximation, for outer or event horizon~(${\cal H}^{+}$)  we get
\begin{eqnarray}
r_{h} \approx r_{+} +\frac{r_{+}^4}{3\left(b^2-2N^2\right)\left(r_{+}-r_{-} \right)}
= r_{+}\left[1+\frac{r_{+}^3}{3\left(b^2-2N^2\right) \left(r_{+}-r_{-}\right)}\right].~\label{eq1.6}
\end{eqnarray}
Similarly, for inner or Cauchy horizon~(${\cal H}^{-}$) we get
\begin{eqnarray}
r_{i} \approx r_{-} +\frac{r_{-}^4}{3\left(b^2-2N^2\right)\left(r_{-}-r_{+} \right)}
= r_{-}\left[1-\frac{r_{-}^3}{3\left(b^2-2N^2\right) \left(r_{+}-r_{-}\right)}\right].~\label{eq1.7}
\end{eqnarray}
Accordingly,
\begin{eqnarray}
r_{h}\,r_{i} \approx r_{+}\,r_{-} \left[1+\frac{r_{+}^3-r_{-}^3}{3\left(b^2-2N^2\right)\left(r_{+}-r_{-}\right)}\right].
~\label{eq1.8}
\end{eqnarray}
and therefore 
\begin{eqnarray}
r_{h}\,r_{i} \approx r_{+}\,r_{-} \left[1+\frac{r_{+}^2+r_{-}^2+r_{+}r_{-}}{3\left(b^2-2N^2\right)}\right].
~\label{eq1.9}
\end{eqnarray}
Now 
\begin{eqnarray}
r_{+}^2+r_{-}^2+r_{+}r_{-}= \frac{4M^2}{\left(1-2\frac{N^2}{b^2}\right)^2}-
\frac{N^2\left(\frac{N^2}{b^2}-1\right)}{\left(1-2\frac{N^2}{b^2}\right)}. ~\label{eq2.0}
\end{eqnarray}
This indicates that 
\begin{eqnarray}
r_{h}\,r_{i} \approx \frac{N^2\left(\frac{N^2}{b^2}-1\right)}{\left(1-2\frac{N^2}{b^2}\right)}
\left[1+\frac{\left\{\frac{4M^2}{\left(1-2\frac{N^2}{b^2}\right)^2}-
\frac{N^2\left(\frac{N^2}{b^2}-1\right)}{\left(1-2\frac{N^2}{b^2}\right)}\right \}}
{3b^2\left(1-2\frac{N^2}{b^2}\right)}\right]. ~\label{eq2.1}
\end{eqnarray}
This can also be written as 
\begin{eqnarray}
r_{h}\,r_{i} \approx \frac{N^2\left(\Lambda N^2-1\right)}{\left(1-2\Lambda N^2\right)}
\left[1+\frac{\Lambda}{3} \frac{\left\{\frac{4M^2}{\left(1-2\Lambda N^2\right)^2}-
\frac{N^2\left(\Lambda  N^2-1\right)}{\left(1-2\Lambda N^2\right)}\right \}}
{\left(1-2\Lambda N^2\right)}+ \mathcal{O}(\Lambda^2) \right]. ~\label{eq2.2}
\end{eqnarray}
Now we need to know the entropy of the BH. It is determined by the following equation 
\begin{eqnarray}
S_{h} &=& \frac{A_{h}}{4} 
      = \frac{1}{4}\int^{2\pi}_0\int^\pi_0\sqrt{g_{\theta\theta}\,g_{\phi\phi}}{\mid }_{r=r_{h}} d\theta\, d\phi 
      = \pi (r_{h}^2+N^2)
\end{eqnarray}
and consequently for ${\cal H}^{-}$, the  entropy becomes 
\begin{eqnarray}
S_{i} &=&  \pi (r_{i}^2+N^2)
\end{eqnarray}
Hence
\begin{align}
S_{h}\, S_{i} = \pi^2 \Biggl[\frac{N^4\left(\Lambda N^2-1\right)^2}{\left(1-2\Lambda N^2\right)^2}
\left\{1+\frac{2}{3}\Lambda \frac{ \left(\frac{4M^2}{(1-2\Lambda N^2)^2}-
\frac{N^2(\Lambda  N^2-1)}{(1-2\Lambda N^2)}\right)}{(1-2\Lambda N^2)}\right\}\
\notag\\
+N^2 \Biggl[\frac{4M^2}{(1-2\Lambda N^2)^2} \left\{1+\frac{2}{3}\Lambda \frac{\left(\frac{4M^2}
{(1-2\Lambda N^2)^2}-\frac{N^2(\Lambda  N^2-1)}{(1-2\Lambda N^2)}\right)}{(1-2\Lambda N^2)}\right\}
\notag\\
-\frac{2N^2\left(\Lambda N^2-1\right)}{\left(1-2\Lambda N^2\right)}
\Biggl\{1+\frac{\Lambda}{3} \frac{\left(\frac{4M^2}{\left(1-2\Lambda N^2\right)^2}-
\frac{N^2\left(\Lambda  N^2-1\right)}{\left(1-2\Lambda N^2\right)}\right)}
{\left(1-2\Lambda N^2\right)}\Biggr\}\Biggr]
\notag\\
+N^4+\mathcal{O}(\Lambda^2) \Biggr]
~\label{eq2.3}   
\end{align} 
This  indicates that the  entropy product is explicitly mass dependent. Taking cognizance of Eq.~(\ref{tnn}), we
find 
\begin{align}
S_{h}\, S_{i} =  \pi^2 \Biggl[\frac{N^2\left(\Lambda N^2-1\right)^2}{\left(1-2\Lambda N^2\right)^2}
\left\{N^2+\frac{2}{3}\Lambda \frac{ \left(\frac{4J_{N}^2}{(1-2\Lambda N^2)^2}-
\frac{N^4(\Lambda  N^2-1)}{(1-2\Lambda N^2)}\right)}{(1-2\Lambda N^2)}\right\}\
\notag\\
+\frac{4J_{N}^2}{(1-2\Lambda N^2)^2} \left\{1+\frac{2}{3}\Lambda \frac{\left(\frac{4J_{N}^2}
{N^2(1-2\Lambda N^2)^2}-\frac{N^2(\Lambda  N^2-1)}{(1-2\Lambda N^2)}\right)}{(1-2\Lambda N^2)}\right\}
\notag\\
-\frac{2N^2\left(\Lambda N^2-1\right)}{\left(1-2\Lambda N^2\right)}
\Biggl\{N^2+\frac{\Lambda}{3} \frac{\left(\frac{4J_{N}^2}{\left(1-2\Lambda N^2\right)^2}-
\frac{N^2\left(\Lambda  N^2-1\right)}{\left(1-2\Lambda N^2\right)}\right)}
{\left(1-2\Lambda N^2\right)}\Biggr\}
\notag\\
+N^4+\mathcal{O}(\Lambda^2) \Biggr]
~\label{eq2.31}   
\end{align} 
which is definitely \emph{mass-independent}.

For our record we also write 
\begin{eqnarray}
r_{h}+r_{i} \approx \frac{2M}{\left(1-2\frac{N^2}{b^2}\right)}
\left[1+\frac{\left\{\frac{4M^2}{\left(1-2\frac{N^2}{b^2}\right)^2}-
\frac{N^2\left(\frac{N^2}{b^2}-1\right)}{\left(1-2\frac{N^2}{b^2}\right)}\right \}}
{3b^2\left(1-2\frac{N^2}{b^2}\right)}\right]. ~\label{eq2.4}      
\end{eqnarray}
which is  again mass dependent. But taking cognizance of Eq.~(\ref{tnn}), we 
see that 
\begin{eqnarray}
r_{h}+r_{i} \approx \frac{2J_{N}}{N\left(1-2\frac{N^2}{b^2}\right)}
\left[1+\frac{\left\{\frac{4J_{N}^2}{N^2\left(1-2\frac{N^2}{b^2}\right)^2}-
\frac{N^2\left(\frac{N^2}{b^2}-1\right)}{\left(1-2\frac{N^2}{b^2}\right)}\right \}}
{3b^2\left(1-2\frac{N^2}{b^2}\right)}\right]. ~\label{eq2.40}      
\end{eqnarray}
this quantity is surely \emph{mass-independent}.

\subsubsection{Cosmological Horizon}
Now we will see where is the cosmological horizon in this perturbative procedure? To do that from the 
exact Killing horizon equation~(\ref{eq1.4})  we have 
\begin{eqnarray}
 r^2 &=& 3(b^2-2N^2) \frac{(r-r_{+})(r-r_{-})}{r^2}  ~\label{eq2.5}
\end{eqnarray}
As a zero order approximation, we get 
\begin{eqnarray}
r_{\Lambda} \approx \sqrt{3} b-\sqrt{3} \frac{N^2}{b}~\label{eq2.6}
\end{eqnarray}
As a first order approximation, we get 
\begin{eqnarray}
 r_{\Lambda} &\approx& \sqrt{3} b\left(1-\frac{N^2}{b^2}\right) 
 \left[\frac{ \left\{\sqrt{3} b\left(1-\frac{N^2}{b^2}\right)-r_{+}\right\}
 \left\{\sqrt{3} b\left(1-\frac{N^2}{b^2}\right)-r_{-}\right\}}
 {\sqrt{3} b\left(1-\frac{N^2}{b^2}\right)}\right]^\frac{1}{2}  \\
 & \approx & \sqrt{3} b\left(1-\frac{N^2}{b^2}\right) \left[1-\frac{r_{+}+r_{-}}
 {2 \sqrt{3} b\left(1-\frac{N^2}{b^2}\right) } \right]\\
 & \approx & \sqrt{3} b\left(1-\frac{N^2}{b^2}\right)-M \left(1+2\frac{N^2}{b^2} \right)
 ~\label{eq2.7}
\end{eqnarray}
Thus the cosmological horizon is situated at
\begin{eqnarray}
r_{\Lambda} \approx \sqrt{3} b-M-\frac{N^2}{b^2}(\sqrt{3}b+2M)~\label{eq2.8}
\end{eqnarray}
It implies that the cosmological horizon depends on both NUT charge and the mass parameter 
including the cosmological constant.

\subsubsection{Virtual Horizon}
Now we will determine the location of virtual horizon~(unphysical) from the exact quartic 
equation. To do that we find from the exact quartic equation
\begin{eqnarray}
r_{V} &=& -\left(r_{h}+r_{i}+r_{\Lambda} \right)                ~\label{eq2.9}
\end{eqnarray}
Thus to a first approximation 
\begin{eqnarray}
r_{V} \approx \sqrt{3} b-M-\frac{N^2}{b^2}(2M-\sqrt{3}b)~\label{eq3.0}
\end{eqnarray}
\subsection{Exact Results}
Which relations are exactly independent of $M$? This is our main aim in this section. To do that 
from exact quartic equation we have 
\begin{eqnarray}
r_{h}\,r_{i}\,r_{\Lambda}\, r_{V} &=& -3N^2(N^2-b^2)  ~\label{eq3.1}
\end{eqnarray}
This can be rewritten in terms of physical horizons
\begin{eqnarray}
r_{h}\,r_{i}\,r_{\Lambda}(r_{h}+r_{i}+r_{\Lambda}) &=& 3N^2(N^2-b^2)                ~\label{eq3.2}
\end{eqnarray}
indicating that it is strictly  independent of $M$. Now we want to write this equation in terms 
entropy to get a relation like $S_{+}S_{-}$. Thus the reduced  entropy could be defined for different 
horizons as 
$$
\tilde{S}_{h}=\frac{S_{h}}{\pi},\,\, \tilde{S}_{i}=\frac{S_{i}}{\pi},\,\, 
\tilde{S}_{\Lambda}=\frac{S_{\Lambda}}{\pi}  
$$
In terms of reduced  entropy one obtains the mass-independent relation 
$$
\left[\sqrt{\tilde{S}_{h} -N^2}+\sqrt{\tilde{S}_{i} -N^2}+\sqrt{\tilde{S}_{\Lambda} -N^2}\right] \times
$$
\begin{eqnarray}
\sqrt{\tilde{S}_{h} -N^2}\,\sqrt{\tilde{S}_{i} -N^2}\,\sqrt{\tilde{S}_{\Lambda} -N^2}
=3N^2(N^2-b^2) ~\label{eq3.3}
\end{eqnarray}
But it not looks like that of $S_{+}S_{-}$. So we take the another condition from quartic equation 
\begin{equation}
\sum_{i>j} r_{i} r_{j} = - 3 (b^2-2N^2). ~\label{eq3.4}
\end{equation}
This implies that 
\begin{equation}
\left(r_{h}  + r_{i} +  r_{\Lambda} \right)\,r_{V}+r_{h} \, \left(r_{i}  + r_{\Lambda}\right) 
+  r_{i}  \,r_{\Lambda} = - 3 (b^2-2N^2), ~\label{eq3.5}
\end{equation}
Moreover,  
\begin{equation}
\left(r_{h}  + r_{i} +  r_{\Lambda}\right)^2-r_{h}\,\left(r_{i}  + r_{\Lambda} \right) -  r_{i}  \, r_{\Lambda} 
= 3 (b^2-2N^2),  ~\label{eq3.6} 
\end{equation}
Furthermore, 
\begin{eqnarray}
r_{h}^2 + r_{i}^2 +  r_{\Lambda}^2 + r_{h} \, r_{i}  +r_{i} \, r_{\Lambda} + r_{\Lambda} \, r_{h} 
= 3 (b^2-2N^2). ~\label{eq3.7}
\end{eqnarray}
This is also explicitly mass-independent. Now in terms of  entropy it can be written as
$$
\left(\tilde{S}_{h}+\tilde{S}_{i}+\tilde{S}_{\Lambda}-3N^2 \right)+\sqrt{\tilde{S}_{h} -N^2}\,\sqrt{\tilde{S}_{i} -N^2}+
$$
\begin{eqnarray}
\sqrt{\tilde{S}_{i} -N^2}\,\sqrt{\tilde{S}_{\Lambda} -N^2} 
+\sqrt{\tilde{S}_{h} -N^2}\,\sqrt{\tilde{S}_{\Lambda} -N^2} &=& 3 (b^2-2N^2).  ~\label{eq3.8}
\end{eqnarray}
Eliminating further one obtains 
\begin{eqnarray}
\frac{r_{h}\,  r_{i} \, r_{\Lambda}\,(r_{h} + r_{i} + r_{\Lambda})}
{r_{h}^2 + r_{i}^2 +  r_{\Lambda}^2+ r_{h}  r_{i}  +r_{i}  \, r_{\Lambda} + r_{\Lambda }\,r_{h}}=
\frac{N^2(N^2-b^2)}{(b^2-2N^2)} ~\label{eq3.9}
\end{eqnarray}
Unquestionably, this is  mass-independent. But it is somehow a complicated function of three 
physical horizon radii. It strictly depends upon the NUT charge and cosmological constat.  

If one can prefer it to write in terms of entropy then the equation would be 
\begin{eqnarray}
f(\tilde{S}_{h},\,\tilde{S}_{i},\,\tilde{S}_{\Lambda}) &=&  
\frac{N^2\left(\frac{N^2}{b^2}-1\right)}{\left(1-2\frac{N^2}{b^2}\right)} ~\label{eq4.0}
\end{eqnarray}
where
$$
f(\tilde{S}_{h},\,\tilde{S}_{i},\,\tilde{S}_{\Lambda}) = 
$$
\begin{eqnarray}
\frac{\left[\sqrt{\tilde{S}_{h} -N^2}+\sqrt{\tilde{S}_{i} -N^2}+\sqrt{\tilde{S}_{\Lambda} -N^2}\right]
\sqrt{\tilde{S}_{h} -N^2}\,\sqrt{\tilde{S}_{i} -N^2}\,\sqrt{\tilde{S}_{\Lambda} -N^2}}
{\left(\tilde{S}_{h}+\tilde{S}_{i}+\tilde{S}_{\Lambda}-3N^2 \right)+
\sqrt{\tilde{S}_{h} -N^2}\,\sqrt{\tilde{S}_{i} -N^2}+\sqrt{\tilde{S}_{i} -N^2}\,\sqrt{\tilde{S}_{\Lambda} -N^2} 
+\sqrt{\tilde{S}_{h} -N^2}\,\sqrt{\tilde{S}_{\Lambda} -N^2}}\nonumber\\                
~\label{eq4.1}
\end{eqnarray}
This is the functional form of three  physical horizon entropy in terms of NUT parameter and 
cosmological constant. Of-course this  is certainly a complicated functional form. 
But it is never near as straight-forward relation like as simple  entropy  products 
of Taub-NUT BH. If we could prefer it to be written in terms of cosmological 
constant then the functional form becomes 
\begin{eqnarray}
f(\tilde{S}_{h},\,\tilde{S}_{i},\,\tilde{S}_{\Lambda}) &=&  
\frac{N^2\left(\Lambda N^2-1\right)}{\left(1-2\Lambda N^2\right)} ~\label{eq4.2}
\end{eqnarray}

\subsection[]{Mass-Independent Relation in terms of two Physical Horizon Areas}
So far, we have derived mass-independent  entropy-functional relation in terms three physical horizons. 
Now in this section for completeness we will derive mass-independent   entropy-functional relation in terms 
two physical horizons. There are three possible situations. First by eliminating virtual horizon and cosmological 
horizon. Second, by eliminating virtual horizon and Cauchy horizon. And finally, by eliminating virtual horizon 
and event horizon. 

First we will eliminate the virtual horizon and cosmological horizon. Then using  Vi\`ete's theorem 
we would obtain following equations from exact quartic 
\begin{eqnarray}
\left(r_{h} + r_{i} \right)^2-r_{h}\, r_{i}+\frac{3N^2(N^2-b^2)}{r_{h}\, r_{i}} &=& 3(b^2-2N^2)   ~\label{eq4.3}
\end{eqnarray}
and 
\begin{eqnarray}
\left(r_{h} + r_{i} \right) \left[r_{h}\, r_{i}+\frac{3N^2(N^2-b^2)}{r_{h}\, r_{i}} \right]
&=& 6 M\,b^2  ~\label{eq4.4}
\end{eqnarray}
Further solving, we can rewrite these equations 
\begin{eqnarray}
r_{h}\, r_{i} &=& \frac{3N^2(N^2-b^2)+r_{h}^2\, r_{i}^2}{3(b^2-2N^2)-(r_{h}^2 + r_{i}^2)} 
   ~\label{eq4.5}
\end{eqnarray}
and 
\begin{eqnarray}
r_{h} + r_{i} &=& \frac{6 M\,b^2}{3(b^2-2N^2)-\left(r_{h}^2+ r_{i}^2\right)}
  ~\label{eq4.6}
\end{eqnarray}
Now we will use these simple formulae to avoid the square root in  entropy 
\begin{eqnarray}
r_{h}^2 + r_{i}^2 &=& \left(r_{h}+ r_{i}\right)^2-2 r_{h}\, r_{i} ~\label{eq4.7}
\end{eqnarray}
and 
\begin{eqnarray}
r_{h}^2\, r_{i}^2=\left(r_{h}\, r_{i} \right)^2   ~\label{eq4.8}
\end{eqnarray}
In terms of  entropy we found that 
$$
\tilde{S}_{h}\,\tilde{S}_{i}\left(\tilde{S}_{h}^2+\tilde{S}_{i}^2+\tilde{S}_{h}\,\tilde{S}_{i} \right) =
N^2 \left(\tilde{S}_{h}+\tilde{S}_{i} \right)\left[\tilde{S}_{h}^2+\tilde{S}_{i}^2-8\tilde{S}_{h}\,\tilde{S}_{i} \right]
+8N^4\left(\tilde{S}_{h}^2+\tilde{S}_{i}^2+\tilde{S}_{h}\,\tilde{S}_{i} \right)
$$
\begin{eqnarray}
+6b^2 \left[\tilde{S}_{h}\,\tilde{S}_{i}\left(\tilde{S}_{h}+\tilde{S}_{i} \right)-
N^2\left(\tilde{S}_{h}^2+\tilde{S}_{i}^2-\tilde{S}_{h}\,\tilde{S}_{i} \right)-
2N^4\left(\tilde{S}_{h}+\tilde{S}_{i} \right)\right]\nonumber\\
-9b^4\left[\tilde{S}_{h}\,\tilde{S}_{i}-N^2\left(\tilde{S}_{h}+\tilde{S}_{i} \right)\right]~\label{eq4.9}
\end{eqnarray}
This is strictly mass-independent. But it is some complicated function of two physical horizons. It can 
also be written as 
$$
\tilde{S}_{h}\,\tilde{S}_{i}\left(\tilde{S}_{h}^2+\tilde{S}_{i}^2+\tilde{S}_{h}\,\tilde{S}_{i} \right) =
N^2 \left(\tilde{S}_{h}+\tilde{S}_{i} \right)\left[\tilde{S}_{h}^2+\tilde{S}_{i}^2-8\tilde{S}_{h}\,\tilde{S}_{i} \right]
+8N^4\left(\tilde{S}_{h}^2+\tilde{S}_{i}^2+\tilde{S}_{h}\,\tilde{S}_{i} \right)
$$
\begin{eqnarray}
+\frac{6}{\Lambda} \left[\tilde{S}_{h}\,\tilde{S}_{i}\left(\tilde{S}_{h}+\tilde{S}_{i} \right)-
N^2\left(\tilde{S}_{h}^2+\tilde{S}_{i}^2-\tilde{S}_{h}\,\tilde{S}_{i} \right)-
2N^4\left(\tilde{S}_{h}+\tilde{S}_{i} \right)\right]\nonumber\\
-\frac{9}{\Lambda^2}\left[\tilde{S}_{h}\,\tilde{S}_{i}-N^2\left(\tilde{S}_{h}+\tilde{S}_{i} \right)\right]~\label{eq5.0}
\end{eqnarray}
When the cosmological constant is switched-off~($\Lambda\rightarrow 0$) one obtains the standard 
result of Taub-NUT BH. When the NUT parameter is switched-off~($N\rightarrow 0$) one obtains the 
mass-independent relation of Kottler spacetime.

For completeness we also note that the  \emph{mass-dependent} relation in terms of two physical 
horizons entropy 
$$
\left(\tilde{S}_{h}+\tilde{S}_{i} \right)
\left[\left(\tilde{S}_{h}^2+\tilde{S}_{i}^2-4\tilde{S}_{h}\,\tilde{S}_{i}\right) 
+12 N^2\left(\tilde{S}_{h}+\tilde{S}_{i} \right)+9b^4-24N^4 \right]
$$
\begin{eqnarray}
-6b^2 \left(\tilde{S}_{h}^2+\tilde{S}_{i}^2+\tilde{S}_{h}\,\tilde{S}_{i}\right) 
&=& 36b^4(M^2+N^2)-72b^2N^4+32N^6                ~\label{eq5.1}
\end{eqnarray}
Whenever we take the new criterion defined in Eq.~(\ref{tnn}), one obtains 
\emph{mass-independent}   entropy functional relation in terms of two physical 
horizons 
$$
N^2\left(\tilde{S}_{h}+\tilde{S}_{i} \right)
\left[\left(\tilde{S}_{h}^2+\tilde{S}_{i}^2-4\tilde{S}_{h}\,\tilde{S}_{i}\right) 
+12 N^2\left(\tilde{S}_{h}+\tilde{S}_{i} \right)+9b^4-24N^4 \right]
$$
\begin{eqnarray}
-6b^2N^2 \left(\tilde{S}_{h}^2+\tilde{S}_{i}^2+\tilde{S}_{h}\,\tilde{S}_{i}\right) 
&=& 36b^4J_{N}^2+4N^4\left(9b^4-18b^2N^2+8N^4\right)    ~\label{eq5.2}
\end{eqnarray}
Remarkably, this is  \emph{mass-independent}. It is possible only due 
to the introduction of conserved charges $J_{N}$ which is quite analogues 
to the Kerr like angular momentum $J=aM$.  This is the \emph{first} time we 
obtain such type of relation.

It may be rewritten as 
\begin{eqnarray}
f(\tilde{S}_{h},\,\tilde{S}_{i}) &=&  \frac{36}{\Lambda^2} J_{N}^2+4N^4\left(\frac{9}{\Lambda^2}
-\frac{18}{\Lambda}N^2+8N^4\right) ~\label{eq5.3}
\end{eqnarray}
where 
$$
f(\tilde{S}_{h},\,\tilde{S}_{i})=N^2\left(\tilde{S}_{h}+\tilde{S}_{i} \right)
\left[\left(\tilde{S}_{h}^2+\tilde{S}_{i}^2-4\tilde{S}_{h}\,\tilde{S}_{i}\right) 
+12 N^2\left(\tilde{S}_{h}+\tilde{S}_{i} \right)+\frac{9}{\Lambda^2}-24N^4 \right]
$$
\begin{eqnarray}
-\frac{6}{\Lambda} N^2 \left(\tilde{S}_{h}^2+\tilde{S}_{i}^2+\tilde{S}_{h}\,\tilde{S}_{i}\right).
~\label{eq5.4}
\end{eqnarray}

\textbf{Case II:}
\vglue 1mm 
When we would eliminate the virtual horizon and the Cauchy horizon then we would obtain 
$$
\left(\tilde{S}_{h}+\tilde{S}_{\Lambda}-2N^2\right) \left[\left(3b^2-2N^2\right)- 
\left( \tilde{S}_{h}+\tilde{S}_{\Lambda}-2N^2\right)\right]^2+
$$
$$
2\left[3N^2(N^2-b^2)+(\tilde{S}_{h}-N^2)(\tilde{S}_{\Lambda}-N^2)\right]\times
\left[3\left(b^2-2N^2 \right)-\left(\tilde{S}_{h}+\tilde{S}_{\Lambda}-2N^2\right)\right] 
$$
\begin{eqnarray}
&=& 36 M^2 b^4. ~\label{eq5.5}
\end{eqnarray}
More explicitly, it can be written as 
$$
\left(\tilde{S}_{h}+\tilde{S}_{\Lambda} \right) \left[\left(\tilde{S}_{h}^2+\tilde{S}_{\Lambda}^2
-4\tilde{S}_{h}\,\tilde{S}_{\Lambda}\right) +12 N^2\left(\tilde{S}_{h}+\tilde{S}_{\Lambda}\right)
+9b^4-24 N^4 \right]
$$
\begin{eqnarray}
-6b^2 \left(\tilde{S}_{h}^2+\tilde{S}_{\Lambda}^2+\tilde{S}_{h}\,\tilde{S}_{\Lambda}\right) 
&=& 36 b^4\, (M^2+N^2)-72b^2\,N^4+32N^6. ~\label{eq5.6}
\end{eqnarray}
Again this is certainly mass-dependent relation of some complicated function of two physical 
horizons  entropy. Remarkably, it turns out to be \emph{mass-independent} when we have taken the 
input of Eq.~(\ref{tnn}). So, it could be written as 
$$
N^2\left(\tilde{S}_{h}+\tilde{S}_{\Lambda} \right)
\left[\left(\tilde{S}_{h}^2+\tilde{S}_{\Lambda}^2-4\tilde{S}_{h}\,\tilde{S}_{\Lambda}\right) 
+12 N^2\left(\tilde{S}_{h}+\tilde{S}_{\Lambda} \right)+9b^4-24N^4 \right]
$$
\begin{eqnarray}
-6b^2 \,N^2 \left(\tilde{S}_{h}^2+\tilde{S}_{\Lambda}^2+\tilde{S}_{h}\,\tilde{S}_{\Lambda}\right) 
&=& 36\,b^4\,J_{N}^2+4N^4\left(9b^4-18b^2N^2+8N^4\right)    ~\label{eq5.7}
\end{eqnarray}
Another \emph{mass-independent}  relation in terms of two physical horizons  entropy namely event 
horizon  entropy and cosmological horizon  entropy is 
$$
\left(\tilde{S}_{h}-N^2 \right)\left(\tilde{S}_{\Lambda}-N^2 \right)
\left[\left(3b^2-2N^2\right)-\left( \tilde{S}_{h}+\tilde{S}_{\Lambda}-2N^2\right)\right]^2
$$
\begin{eqnarray}
&=& \left[3N^2(N^2-b^2)+(\tilde{S}_{h}-N^2)(\tilde{S}_{\Lambda}-N^2)\right]^2.~\label{eq5.8}
\end{eqnarray}
More explicitly,
$$
\tilde{S}_{h}\,\tilde{S}_{\Lambda}\left(\tilde{S}_{h}^2+\tilde{S}_{\Lambda}^2+\tilde{S}_{h}\,\tilde{S}_{\Lambda} \right) =
N^2 \left(\tilde{S}_{h}+\tilde{S}_{\Lambda} \right)\left[\tilde{S}_{h}^2+\tilde{S}_{\Lambda}^2
-8\tilde{S}_{h}\,\tilde{S}_{\Lambda} \right]
+8N^4\left(\tilde{S}_{h}^2+\tilde{S}_{\Lambda}^2+\tilde{S}_{h}\,\tilde{S}_{\Lambda} \right)
$$
\begin{eqnarray}
+6 b^2 \left[\tilde{S}_{h}\,\tilde{S}_{\Lambda}\left(\tilde{S}_{h}+\tilde{S}_{\Lambda} \right)-
N^2\left(\tilde{S}_{h}^2+\tilde{S}_{\Lambda}^2-\tilde{S}_{h}\,\tilde{S}_{\Lambda} \right)-
2N^4\left(\tilde{S}_{h}+\tilde{S}_{\Lambda} \right)\right]\nonumber\\
-9b^4\left[\tilde{S}_{h}\,\tilde{S}_{\Lambda}-N^2\left(\tilde{S}_{h}
+\tilde{S}_{\Lambda} \right)\right]. ~\label{eq5.9}
\end{eqnarray}
Analoguosly, by eliminating virtual horizon and event horizon one could obtain 
mass-independent relations in terms of cosmological horizon and Cauchy horizon. 
Now we will turn to the AdS case.

\section{\label{tnads}~Taub-NUT--anti-de~Sitter BHs}
Let us put in Eq.~(\ref{eq1.1})
$$
\Lambda=-\frac{1}{|b|^2},
$$
then the relevent quartic equation will be 
\begin{eqnarray}
r^4+3\left(2N^2+|b|^2\right)r^2-6\,M\, r\,|b|^2-3\,N^2\,|b|^2-3\,N^4 &=& 0  ~\label{e3.1}
\end{eqnarray}
This quartic equation has four roots. Among four roots, two roots are complex and two roots 
are real. Complex roots correspond to two virtual horizons, $r_{V}^{\pm}$. These horizons are 
unphysical. While real roots correspond to two physical horizons: an event~(or outer) horizon 
$r_{h}$ and a Cauchy horizon or inner horizon $r_{i}$. 

\subsection{Perturbative Results}
Proceeding analogously to a first approximation for ${\cal H}^{+}$,  we have
\begin{eqnarray}
r_{h} \approx r_{+} -\frac{r_{+}^4}{3\left(|b|^2+2N^2\right)\left(r_{+}-r_{-} \right)}
= r_{+}\left[1-\frac{r_{+}^3}{3\left(|b|^2+2N^2\right) \left(r_{+}-r_{-}\right)}\right]. ~\label{e3.2}
\end{eqnarray}
Similarly, for ${\cal H}^{-}$ we find
\begin{eqnarray}
r_{i} \approx r_{-} +\frac{r_{-}^4}{3\left(|b|^2+2N^2\right)\left(r_{+}-r_{-}\right)}
= r_{-} \left[1+\frac{r_{-}^3}{3\left(|b|^2+2N^2\right)\left(r_{+}-r_{-}\right)}\right].~\label{e3.3}
\end{eqnarray}
Resultantly,
\begin{eqnarray}
r_{h}\,r_{i} \approx r_{+}\,r_{-} \left[1-\frac{r_{+}^3-r_{-}^3}{3\left(|b|^2+2N^2\right)
\left(r_{+}-r_{-}\right)}\right].~\label{e3.4}
\end{eqnarray}
and thus 
\begin{eqnarray}
r_{h}\,r_{i} \approx r_{+}\,r_{-} \left[1-\frac{r_{+}^2+r_{-}^2+r_{+}r_{-}}
{3\left(|b|^2+2N^2\right)}\right]. ~\label{e3.5}
\end{eqnarray}
Since 
\begin{eqnarray}
r_{+}^2+r_{-}^2+r_{+}r_{-}= \frac{4M^2}{\left(1+2\frac{N^2}{|b|^2}\right)^2}+
\frac{N^2\left(\frac{N^2}{|b|^2}+1\right)}{\left(1+2\frac{N^2}{|b|^2}\right)}. 
~\label{e3.6}
\end{eqnarray}
So, 
\begin{eqnarray}
r_{h}\,r_{i} \approx -\frac{N^2\left(\frac{N^2}{|b|^2}+1\right)}
{\left(1+2\frac{N^2}{|b|^2}\right)} 
\left[1-\frac{\left\{\frac{4M^2}{\left(1+2\frac{N^2}{|b|^2}\right)^2}+
\frac{N^2\left(\frac{N^2}{|b|^2}+1\right)}{\left(1+2\frac{N^2}{|b|^2}\right)}\right\}}
{3|b|^2\left(1+2\frac{N^2}{|b|^2}\right)}\right]. ~\label{e3.7}
\end{eqnarray}
If we have considered the case $\Lambda<0$ then we are in a situation where the cosmological 
horizon does not exist but have asymptotically AdS spacetime. Only two physical horizons 
${\cal H}^{\pm}$ are exist. Therefore
\begin{eqnarray}
r_{h}\,r_{i} \approx -\frac{N^2\left(|\Lambda| N^2+1\right)}{\left(1+2|\Lambda| N^2\right)}
\left[1-\frac{|\Lambda|}{3} \frac{\left\{\frac{4M^2}{\left(1+2|\Lambda| N^2\right)^2}+
\frac{N^2\left(|\Lambda|  N^2+1\right)}{\left(1+2|\Lambda| N^2\right)}\right\}}
{\left(1+2|\Lambda| N^2\right)}+ \mathcal{O}(\Lambda^2) \right]. ~\label{e3.8}
\end{eqnarray}
This suggests that for \emph{either} sign of the cosmological constant one has 
\begin{eqnarray}
r_{h}\,r_{i} \approx -\frac{N^2\left(\Lambda N^2+1\right)}{\left(1+2\Lambda N^2\right)}
\left[1+\frac{\Lambda}{3} \frac{\left\{\frac{4M^2}{\left(1+2 \Lambda N^2\right)^2}+
\frac{N^2\left(\Lambda  N^2+1\right)}{\left(1+2 \Lambda N^2\right)}\right\}}
{\left(1+2\Lambda N^2\right)}+ \mathcal{O}(\Lambda^2) \right]. ~\label{e3.9}
\end{eqnarray}
Note that it is evidently mass-dependent. But taking the cognizance of Eq.~(\ref{tnn}), one obtains 
\begin{eqnarray}
r_{h}\,r_{i} \approx -\frac{\left(\Lambda N^2+1\right)}{\left(1+2\Lambda N^2\right)}
\left[N^2+\frac{\Lambda}{3} \frac{\left\{\frac{4J_{N}^2}{\left(1+2 \Lambda N^2\right)^2}+
\frac{N^4\left(\Lambda  N^2+1\right)}{\left(1+2 \Lambda N^2\right)}\right\}}
{\left(1+2\Lambda N^2\right)}+ \mathcal{O}(\Lambda^2) \right]. ~\label{e3.90}
\end{eqnarray}
Now it is definitely \emph{mass-independent}.

\subsection{Exact Results}
The exact results could be obtain by taking the combinations of various roots. Using 
Vi\`ete's theorem  we have 
\begin{eqnarray}
r_{h}+r_{i}+r_{V}^{+}+ r_{V}^{-} &=& 0,  \\
r_{h}\,r_{i}+\left(r_{h}+r_{i}\right)\left(r_{V}^{+}+ r_{V}^{-}\right)+r_{V}^{+}\, r_{V}^{-} 
&=& 3\left(|b|^2+2N^2\right),\\
r_{h}\,r_{i}\left(r_{V}^{+}+ r_{V}^{-}\right)+r_{V}^{+}\, r_{V}^{-}\left(r_{h}+r_{i}\right) &=& 
6 M\, |b|^2,\\
r_{h}\,r_{i}\,r_{V}^{+}\, r_{V}^{-} &=& -3\,N^2\left(|b|^2+N^2\right). ~\label{e4.1}
\end{eqnarray}
Since we have earlier mentioned that there exists two physical horizons thus we 
are eliminating the virtual horizons due to the unphysical nature. So, we obtain 
\begin{eqnarray}
r_{h}\, r_{i} &=& -\frac{3N^2(N^2+|b|^2)+r_{h}^2\, r_{i}^2}{3(|b|^2+2N^2)+\left(r_{h}^2 + r_{i}^2\right)} 
   ~\label{e4.2}
\end{eqnarray}
and 
\begin{eqnarray}
r_{h} + r_{i} &=& \frac{6 M\,|b|^2}{3(|b|^2+2N^2)+\left(r_{h}^2+ r_{i}^2\right)}
  ~\label{e4.3}
\end{eqnarray}
As previously, to avoid the square root in  entropy formula we have to use the formulae of 
Eq.~(\ref{eq4.7}) and Eq.~(\ref{eq4.8}). So, in terms of area (or entropy) of two physical horizons 
we found that mass-independent relation 
$$
\left(\tilde{S}_{h}-N^2 \right)\left(\tilde{S}_{i}-N^2 \right)
\left[3|b|^2+4N^2+\tilde{S}_{h}+\tilde{S}_{i}\right]^2
$$
\begin{eqnarray}
&=& \left[3N^2(N^2+|b|^2)+(\tilde{S}_{h}-N^2)(\tilde{S}_{i}-N^2)\right]^2.~\label{e4.4}
\end{eqnarray}
Also we find another important equation  
$$
\left(\tilde{S}_{h}+\tilde{S}_{i}-2N^2 \right)
\left[3|b|^2+4N^2+\tilde{S}_{h}+\tilde{S}_{i}\right]^2
$$
$$
-2\left[3N^2(N^2+|b|^2)+(\tilde{S}_{h}-N^2)(\tilde{S}_{i}-N^2)\right]
\left[3|b|^2+4N^2+\tilde{S}_{h}+\tilde{S}_{i}\right]
$$
\begin{eqnarray}
&=& 36\, M^2 |b|^4.~\label{e4.5}
\end{eqnarray}
which is strictly mass-dependent.  Taking cognizance of Eq.~(\ref{tnn}), we can make it 
\emph{mass-independent}. 
$$
N^2\left(\tilde{S}_{h}+\tilde{S}_{i}-2N^2 \right)
\left[3|b|^2+4N^2+\tilde{S}_{h}+\tilde{S}_{i}\right]^2
$$
$$
-2N^2 \left[3N^2(N^2+|b|^2)+(\tilde{S}_{h}-N^2)(\tilde{S}_{i}-N^2)\right]
\left[3|b|^2+4N^2+\tilde{S}_{h}+\tilde{S}_{i}\right]
$$
\begin{eqnarray}
&=& 36\, J_{N}^2 |b|^4.~\label{e4.6}
\end{eqnarray}
This is in fact strictly \emph{mass-independent}. This is possible only due to the factor 
$J_{N}$, a new conserved charges in NUT class of geometry. All above computations demonstrate 
that the  entropy products for NUT class of BHs in AdS space are \emph{generic} as 
well as \emph{universal}.

\section{\label{con} Conclusions}~
It has been argued that a generic four dimensional Taub-NUT BH should be completely
described in terms of three or four different types of thermodynamic hairs. They should be defined 
as the Komar mass~($M=m$), the angular momentum~($J_{n}=mn$), the gravitomagnetic charge ($N=n$), 
the dual~(magnetic) mass $(\tilde{M}=n)$. Under these circumtances, we examined the generic feature of 
area~(or entropy) products of NUT class of BHs in AdS space. 

Particularly, we considered Taub-NUT-de~Sitter BHs and Taub-NUT--anti-de~Sitter BHs. 
First, we  calculated perturbatively the  entropy products in terms of event horizon and Cauchy horizon. 
We showed that the products have mass-independent feature. Next we computed the  entropy products in 
terms of three physical horizons by direct method and proved that the functional form in terms of 
three physical horizons is independent of the mass parameter but it depends on only the NUT 
parameter and cosmological constant.  Then we computed the mass-independent relation in terms of two
physical horizons. 

The analysis that we perfomed  in this work indicates the mass-independent features of NUT class of BHs in AdS space  
are interestingly shown to be \emph{generic as well as universal} which was earlier stated in the literature 
as ``mass-dependent'' characteristics for Schwarzschild--de~Sitter BH and 
Schwarzschild-anti-de~Sitter BH in $3+1$ dimensions. The mass-independent feature or universality 
comes solely  due to the presence of new global conserved charges  $J_{N}=MN$ which is closely analogue to 
the  Kerr like  angular momentum $J=aM$. 

Finally, the Visser's conjectured ``the mass-independence often fails once a cosmological constant is added'' 
is \emph{violated} in the present work.  The generic feature of area ~(or entropy) products that we have examined 
for NUT class of BHs in AdS space has deep significance for further possible
explanation of the microscopic nature of BHs in ($3+1$) dimensions.

Recently, we defined a entropy product function~(EPF)~\cite{pp23} for NUT class of BHs 
and we found that there is an remarkable relation exist between the central charges 
and the EPF as 
$$
c=6\left(\frac{\partial {\cal F}}{\partial {\cal N}_{i}}\right)
$$ 
where ${\cal F}= \frac{{\cal S}_{h}{\cal S}_{i}}{4\pi^2}$ is EPF and ${\cal N}_{i}$ is one of the 
integer valued  charges i.e. the NUT charges~($N$) or any new conserved charges~($J_{N}$). 
It would be an interesting topic of research if we could answer that ``Does it work for NUT class of 
BHs in AdS spacetime like Eq.~(\ref{eq2.31})?. Because the entropy product is mass independent for 
this case. Thus from Eq.~(\ref{eq2.31}), we can easily compute the central charges for NUT 
picture, $J_{N}$ picture and cosmological picture i.e. $\Lambda$ picture. 

\vfill
\emph{Data Availability Statement:}\\
All datasets generated and analysed during this study are available in the manuscript.

\vfill
\emph{Conflict of Interest Statement:}\\
The author here by declare that there is no conflict of interest

\bibliography{sn-bibliography}

\end{document}